\documentclass{article}
\usepackage{amsmath, amssymb}
\usepackage{fullpage}
\usepackage{todonotes}
\usepackage[inkscapelatex=false]{svg}
\usepackage{authblk}
\usepackage{listings} 
\usepackage{dsfont}
\usepackage{url}
\usepackage[numbers,sort&compress]{natbib}

\usepackage[normalem]{ulem}

\begin{document}

\setlength{\parindent}{0pt}
\setlength{\parskip}{1em}

\onecolumn

\renewcommand{\abstractname}{}

\title{Cell--cell adhesion cannot sustain extended follower streams in \\ a minimal non-local model of leader--follower migration}
\newcommand\CoAuthorMark{\footnotemark[\arabic{footnote}]}
\author[1]{Thomas Jun Jewell\thanks{These authors contributed equally and share first authorship.}}
\author[1]{Samuel W.S. Johnson\protect\CoAuthorMark}
\author[1]{Ruth E. Baker\thanks{These authors contributed equally and share senior authorship.}}
\author[1]{Philip K. Maini\protect\CoAuthorMark}
\affil[1]{Wolfson Centre for Mathematical Biology, Mathematical Institute, University of Oxford, Oxford, United Kingdom}

\date{}
\maketitle

\linespread{2}\selectfont

\begin{abstract}
\noindent Cell--cell adhesion is widely hypothesised to maintain cohesion within the long streams of follower cells that trail leader subpopulations during collective migration, including in neural crest cell migration, angiogenesis, and cancer cell invasion. Mathematically, non-local advection--diffusion equations provide the canonical continuum framework within which to study such adhesive cell--cell interactions. Here, we study a minimal model of leader--follower migration within this framework, in which leaders migrate at constant velocity while followers are attracted to leaders and to one another over a finite spatial interaction range. Numerical simulations reveal that, although the model can maintain small cohorts of travelling follower cells, the size of these cohorts is limited by the adhesive interaction lengthscale, and is far below what is needed to reproduce the extended streams observed \textit{in vivo}. This points to a structural limitation of the standard non-local adhesion formulation and highlights the need for the development of extended continuum models capable of sustaining long, coherent migratory streams through purely mass-conserving collective cell movement.
\end{abstract}

\newpage

\subsection*{Introduction}

Collective motion is widespread across biology, from flocking birds \citep{farine2022collective} and schooling fish \citep{lopez2012behavioural}, to the coordinated migration of cells \textit{in vivo} \citep{rorth2009collective}. A recurring feature of such systems is \textit{leader--follower} dynamics, in which a small subset of individuals (leaders) directs the movement of a much larger population (followers). In multicellular systems, leader--follower migration underpins key processes including embryonic development \citep{scarpa2016collective}, wound healing \citep{li2013collective}, and cancer invasion \citep{friedl2004collective}. \textit{In vivo}, leader cells migrate in response to external guidance cues such as chemotactic, durotactic, or haptotactic gradients \citep{sengupta2021principles}, while follower cells coordinate their motion through interactions with neighbouring cells \citep{mishra2019cell}. This leads to the formation of large, connected streams of follower cells trailing leader subpopulations. A canonical example of this phenomenon occurs in chick cranial neural crest migration, where leader cells respond to gradients in vascular endothelial growth factor and follower cells migrate collectively behind them, with cell--cell adhesion hypothesised to maintain cohesion within the stream\,\citep{kulesa2010cranial}. From a mathematical modelling perspective, such spatially extended streams present a challenge, as they persist over distances far greater than the range of direct cell–cell interactions.

Spatially extended migratory fronts arise naturally in mathematical models that include cell proliferation. Classical reaction–diffusion models such as the Fisher–KPP equation produce travelling waves in which cell density behind the advancing front remains high \citep{murray_book2004_FisherKPP, simpson2024_FisherKPP}. Similar mechanisms have been proposed in the context of neural crest migration in the gut. For example, continuum partial differential equation models \citep{simpson2006looking, simpson2007cell} have predicted that proliferation at the migratory wavefront of neural crest streams drives invasion of the developing gut, a prediction later supported by tissue graft experiments in quail. However, in other systems, including the chick cranial neural crest, proliferation during migration is negligible \citep{ridenour2014neural}. In such settings, follower streams must be driven by cell--cell interactions that induce movement, rather than by proliferation. Chemotaxis-based continuum models \citep{mehmetUccar2025_kellerSegel} have been proposed to explain follower migration in such contexts, but typically assume that cells can sense arbitrarily small chemical gradients. Explaining how spatially extended streams emerge without proliferation or long-range gradient sensing, therefore, presents a significant modelling challenge.

Cell--cell adhesion is frequently proposed as a mechanism for coordinating follower motion in such systems\,\citep{kulesa2010cranial}. A canonical and widely used continuum framework for modelling such interactions utilises non-local advection–diffusion equations, also known as non-local aggregation equations, to represent the evolution of interacting cell populations \citep{armstrong2006continuum, carrillo2019_adhesion_nonlocal, painter2024_nonlocalreview}. In this framework, the direction of cell migration depends on cell-density gradients sensed over a finite range, reflecting the inherently non-local nature of cell–cell interactions. This formulation captures the fact that cells can attract or respond to neighbouring cells across finite distances through adhesive, chemical, or mechanical cues. Such models have proven remarkably versatile in the study of collective cell behaviours, reproducing a wide repertoire of multicellular patterning \citep{carrillo2019_adhesion_nonlocal, giunta2025phylogeny}.

A notable example illustrating the power of this framework is provided by Carrillo et al.~\citep{carrillo2019_adhesion_nonlocal}, who showed that this class of models can reproduce, both qualitatively and quantitatively, the sorting and mixing of cell sub-populations driven by differential adhesion, as measured experimentally by Katsunuma et al.~\citep{katsunuma2016_experiments_that_jose_reproduced}. More broadly, non-local advection--diffusion models have been applied across developmental and pathological contexts, providing mechanistic explanations for cell aggregation during development \citep{yu_dagmar_2025_development, jewell2023_dimension, painter2015nonlocal} and forming the basis for representing cell--cell adhesion in a range of processes including cancer invasion \citep{domschke2014_cancer_invasion, gerisch2008_cancer_invasion, painter2010_cancer_invasion}, vasculogenesis\,\citep{VILLA2022_vasculogenesis}, and skeletal morphogenesis\,\citep{glimm2014_skeleton_morphogenesis}. For an extensive review of these models, see \citep{painter2024_nonlocalreview}.

Accordingly, the non-local advection--diffusion framework provides a canonical continuum description of cell--cell adhesion and therefore a natural setting in which to study leader--follower migration. However, an unresolved question in this context is whether models based solely on co-attraction and adhesion between proximal cells can sustain streams of follower cells that are both motile and span distances far greater than the typical range of cell--cell interactions. Previous analyses have shown that such models can produce stationary clusters whose extent exceeds the interaction lengthscale of cells \citep{potts2024_nonlocal_long_time_steady_states, jewell2025chase_run}, or travelling pulses whose size remains limited by that length scale \citep{carrillo2018zoology, painter2024chase_run, jewell2025chase_run}, yet long, cohesive streams of migrating followers have not been reported.

\begin{figure}
    \centering
    \includegraphics[width=\linewidth]{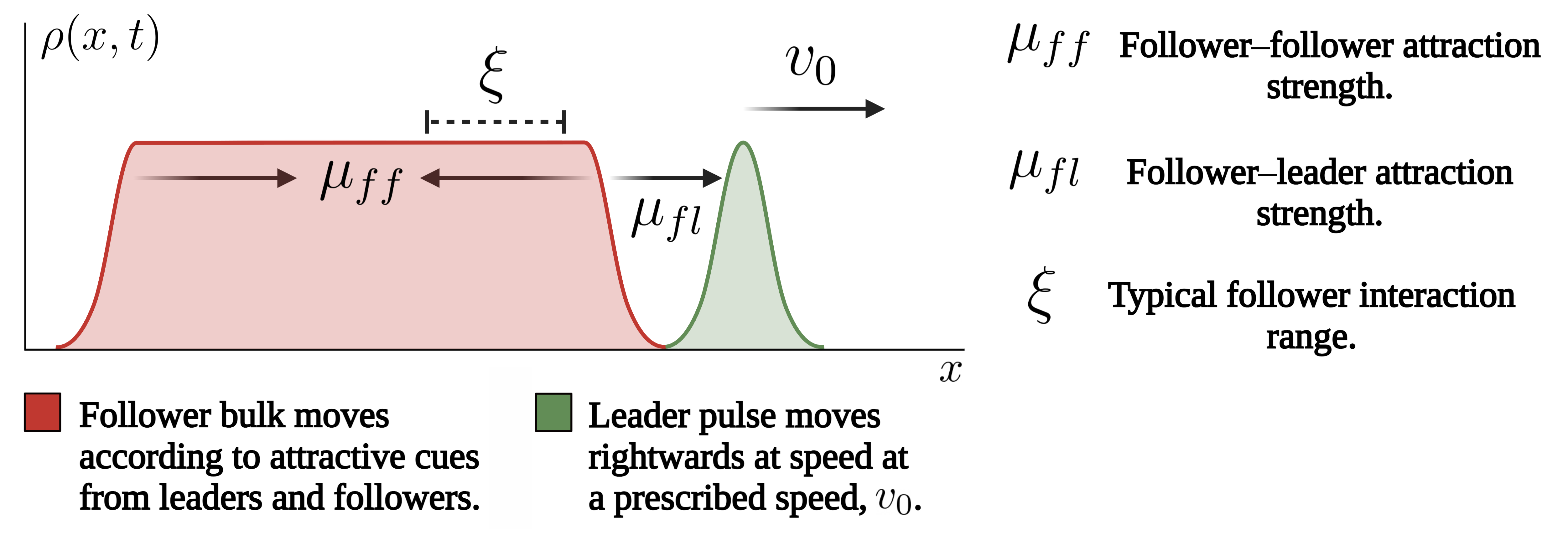}
    \caption{Mathematical model schematic. The model represents leader and follower populations as continuous densities in a one-dimensional domain. Leader cells migrate in the positive $x$-direction with a constant speed, $v_0$, representing migration according to gradients in external cues. Follower cells move according to an integro-partial differential equation, with terms representing diffusion and attractive cues from leader cells and other follower cells. Follower--follower and follower--leader interaction strengths are modulated by the parameters $\mu_{ff}$ and $\mu_{fl}$, respectively, while the typical length over which follower--follower and follower--leader interactions occur is represented by the parameter $\xi$.}
    \label{fig:introductionFigure}
\end{figure}

To address this question, we study a minimal mathematical model of leader–follower migration formulated within the non-local advection–diffusion framework. Leader cells are prescribed to migrate at constant velocity in a one-dimensional domain, representing directed motion in response to external cues, while follower cells move according to short-ranged attractive interactions with both leaders and other followers (Figure \ref{fig:introductionFigure}).

\subsection*{Methods}
\label{methods}

We study a minimal continuum model of leader–follower migration on a one-dimensional domain of $2000\,\mu\rm{m}$ in length. 
The model comprises two cell populations, leaders and followers, each represented as continuously varying densities along the length of the domain. 
Leaders and followers each evolve in time according to a system of coupled partial differential equations (PDEs). 
The leader PDE prescribes movement in the positive $x$-direction at a fixed speed, $v_0$, representing migration according to local micro-environmental cues, in processes such as chemotaxis.
Conversely, the follower PDE contains terms representing movement according to random diffusive motion, reflecting undirected motility with diffusivity $D_f$, and short-ranged, non-local attraction to both leaders and other followers, representing the adhesive and co-attractive interactions commonly cited as key mechanisms for follower cell migration. This effectively extends previous work by Potts\,\,\citep{potts2025aggregation}, which examined population organisation around a stationary resource, to the case of a moving resource, here represented by the leaders. Follower--follower and follower--leader interactions are parameterised by the (positive) interaction strength coefficients $\mu_{ff}$ and $\mu_{fl}$, respectively, and by an interaction lengthscale, $\xi$, that determines the typical distance over which cells can sense and interact with one another. 
Here, we take $\xi$ to be the same for follower--follower and follower--leader interactions, as such interactions are often mediated by intrinsic properties of follower cells, such as the length of actin-based protrusions extended at their membrane, or the density of ligand receptors along their membrane.
This assumption has also been made in a previous model of non-local leader--follower interactions \citep{painter2024chase_run}.
Follower motion is also subject to a volume-filling constraint, which prevents overcrowding when the combined leader–follower density approaches a maximum density, $\kappa$. 

Under these assumptions, the evolution of the leader and follower cell densities, $\rho_l(x,t)$ and $\rho_f(x,t)$, in the domain $x \in [0,L]$ where $L=2000\,\mu\mathrm{m}$, is governed by the PDE
\begin{equation}
\begin{split}
   &\frac{\partial \rho_{l}}{\partial t} = -v_0\,\frac{\partial \rho_{l}}{\partial x}, \\[6pt]
&\frac{\partial \rho_{f}}{\partial t} = D_{f}\,\frac{\partial^2 \rho_{f}}{\partial x^2}
\;-\;\frac{\partial}{\partial x}\!\left(\rho_{f}\,(\kappa-\rho_l-\rho_f)\,\frac{\partial C}{\partial x}\right),
\label{eq:main_PDE} 
\end{split}
\end{equation}

where $C$ is the non-local attraction potential, incorporating follower--follower and leader--follower interactions,
\begin{equation}
C(x,t) = \mu_{ff}\!\int_{0}^{L} K_{\xi}(x-y)\,\rho_{f}(y,t)\,\mathrm{d}y
+ \mu_{fl}\!\int_{0}^{L} K_{\xi}(x-y)\,\rho_{l}(y,t)\,\mathrm{d}y.
\label{eq:attraction}
\end{equation}
The kernel, $K_{\xi}(x)$, specifies how the strength of co-attraction or adhesion changes with the distance between cells, and is taken as an even, non-negative Gaussian function:
\begin{equation}
K_{\xi}(x) \;=\;
\begin{cases}
\alpha(\xi)
\exp\!\left(-\dfrac{x^{2}}{2 \xi^{2}}\right),
& |x|\leq 3\xi, \\[10pt]
0, & |x|>3\xi , 
\end{cases}
\label{eq:kernel_main}
\end{equation}
truncated at $3\xi$ to ensure compact support while retaining over 99\% of the Gaussian mass. The factor $\alpha(\xi)$ ensures that 
\begin{equation}
\int_{\mathbb{R}} K_{\xi}(x)\,\mathrm{d}x = \xi, 
\end{equation}
such that, unlike many continuum representations of non-local adhesion, we do not normalise the kernel $K_{\xi}(x)$ independently of $\xi$. Instead, we allow the total mass of the kernel to scale proportionally with $\xi$, reflecting the modelling assumption that increasing the interaction range introduces additional interactions rather than diluting the same interactions over a larger distance. Simulations using alternative interaction kernels, including top-hat and Hookean spring-type functions, display qualitatively similar behaviour, suggesting that the results are qualitatively robust to the choice of kernel. Further details of the model and the numerical scheme with which these equations are solved can be found in Appendix~A with the corresponding model code available at \url{https://github.com/SWSJChCh/continuumLeaderFollower}. 

At the beginning of each simulation, leaders are initialised as a narrow cluster positioned closely ahead of a bulk of follower cells that extend over several hundred micrometres (Figure \ref{fig:introductionFigure}). 
Leaders then advance with a constant velocity of $1\,\mu\rm{m}\,\rm{min}^{-1}$ for a period of $12$ hours, while followers disperse, interact, and reorganise according to Equation\,\eqref{eq:main_PDE}. 
Parameter values of the model are chosen to represent typical cranial neural crest cell dynamics \textit{in vivo}. Follower diffusivity is fixed at $D_f = 10\,\mu\mathrm{m}^2$ min$^{-1}$, leader speed at $v_0 = 1\,\mu\mathrm{m}$ min$^{-1}$, and follower interaction lengthscales $\xi$ range from approximately one to ten cell lengths ($10\,\mu$m to $100\,\mu$m). 
The interaction strengths $\mu_{ff}$ and $\mu_{fl}$ are varied systematically to explore the extent to which follower--follower and follower--leader adhesion and co-attraction facilitate the formation of extended streams of follower cells migrating as a collective behind leader cells.

\begin{figure}
    \centering
    \includegraphics[width=0.7\linewidth]{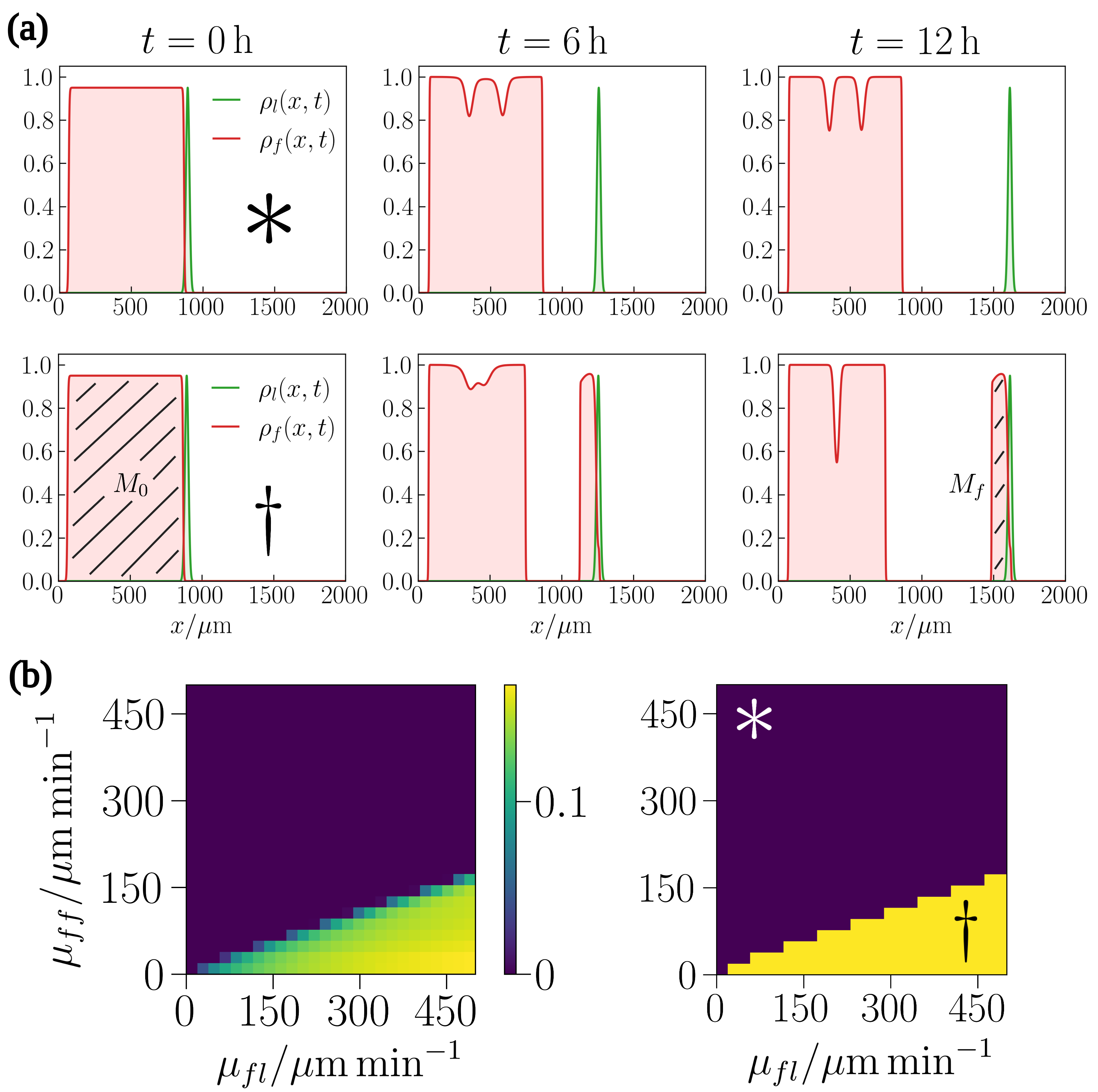}
    \caption{\textbf{(a)} Model simulations for $(\xi, \mu_{ff}, \mu_{fl}) = (50\,\mu\rm{m}, 40\,\mu\rm{m}\,\rm{min}^{-1}, 40\,\mu\rm{m}\,\rm{min}^{-1})$ (upper) and $(50\,\mu\rm{m}, 40\,\mu\rm{m}\,\rm{min}^{-1}, 400\,\mu\rm{m}\,\rm{min}^{-1})$ (lower), shown at $0\, \mathrm{h},\ 6\, \mathrm{h},$ and $12\, \mathrm{h}$. In the upper panel, follower–follower and leader–follower attraction strengths are equal. Under these conditions, follower cells (red) remain tightly-bound, and the attraction exerted by leader cells (green) is insufficient to detach followers from the initial cluster. In the lower panel, follower–leader attraction is increased to ten times the follower–follower attraction. As a result, followers located within the initial interaction range of leaders are rapidly advected towards leaders, while those outside of this range remain in the bulk due to comparatively low follower–follower attraction strength. In the lower panel, $M_0$ denotes the initial mass of the follower cluster and $M_f$, the final mass advected with the leaders. \textbf{(b)} Heatmaps showing $M_f/M_0$ for $\xi=50\,\mu\rm{m}$ across a range of follower--follower and follower--leader interaction strengths. The left heatmap reveals two distinct regimes: ($*$) a regime in which follower–follower attraction is sufficiently strong to maintain cluster integrity but followers do not migrate, and ($\dag$) a regime in which strong leader–follower attraction advects only those followers within the leaders’ initial interaction radius. The right heatmap provides a binary classification of these regimes.}
    \label{fig:figure2}
\end{figure}

To quantify the extent to which each parameter set, $(\xi, \mu_{ff}, \mu_{fl})$, facilitates the formation of follower streams, we compare the mass of the main contiguous region of follower density directly behind the leaders after migration ends ($M_f$) to the initial mass of followers in the domain ($M_0$) (Figure \ref{fig:figure2}(a)). This ratio provides a measure of leader--follower migratory efficiency, quantifying the fraction of follower mass that is successfully transported across the tissue segment by leaders. Our analysis compares follower stream sizes across parameter sets and thus investigates whether the model considered here is sufficient to reproduce the long streams of followers observed \textit{in vivo}. Systematic explorations of parameter space have been used to classify pattern-forming behaviour in related non-local advection–diffusion models \citep{giunta2025phylogeny}.

\begin{figure}
    \centering
    \includegraphics[width=\linewidth]{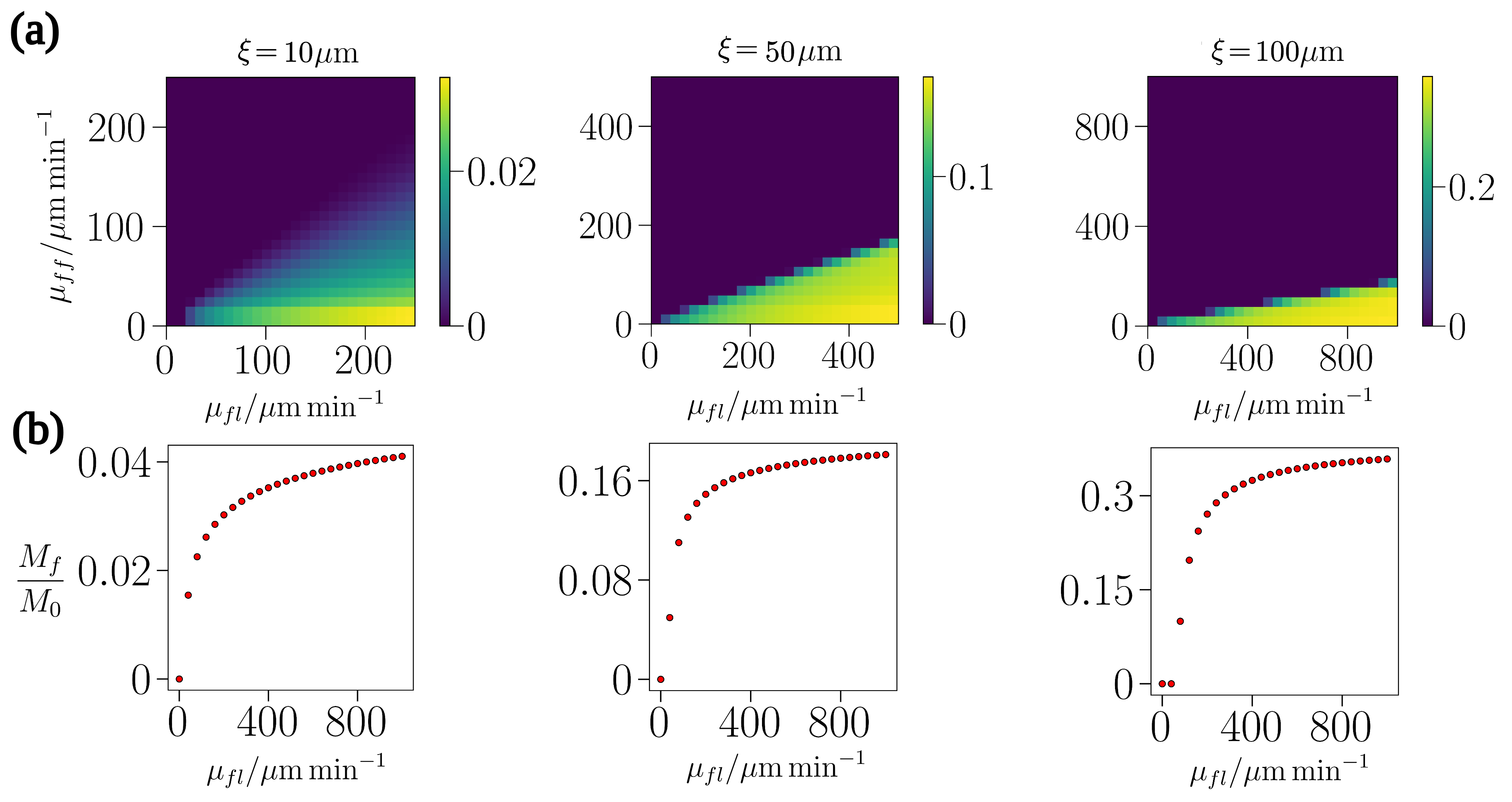}
    \caption{\textbf{(a)} Heatmaps of $M_f/M_0$ for $\xi=10\,\mu\rm{m}, 50\,\mu\rm{m}, 100\,\mu\rm{m}$ for a range of interaction strengths. Heatmaps show the existence of two regimes in $(\mu_{ff}, \mu_{fl})$ space; a regime in which follower–follower attraction is sufficiently strong to maintain cluster integrity but followers do not migrate, and a regime in which strong leader–follower attraction advects only those followers within the leaders’ initial interaction radius. \textbf{(b)} Line plots of $M_f/M_0$ as a function of $\mu_{fl}$ for fixed $\mu_{ff}=10\,\mu\rm{m}\,\rm{min}^{-1}$. Line plots show a gradual plateau in $M_f/M_0$ as $\mu_{fl}$ is increased, indicating a fundamental limit on the degree to which long follower chains may be created.}
    \label{fig:figure3}
\end{figure}

\section*{Results}

We systematically explored the parameter space defined by the interaction strengths $(\mu_{ff}, \mu_{fl})$ and interaction lengthscale, $\xi$, quantifying migratory success by the ratio $M_f/M_0$ of follower mass transported along the domain with leaders relative to the total follower mass. Parameter sweeps were performed over wide ranges of $\mu_{ff}$ and $\mu_{fl}$ for $\xi = 10, 50, 100\,\mu$m, representing characteristic interaction lengthscales of approximately one, five, and ten cell lengths, respectively (with interactions 
extending to $3\xi$). 

For all values of $\xi$ considered, two distinct dynamical regimes consistently emerged (Figure \ref{fig:figure3}). In the first regime, follower–follower attraction, $\mu_{ff}$, is sufficiently strong, such that attraction from leader cells is unable to detach follower cells from the initial bulk, and thus, the entire follower mass remains in place. An example simulation in this regime is shown in the upper panels of Figure \ref{fig:figure2}(a) and corresponds to the region denoted by $*$ in the binary heatmap shown in Figure \ref{fig:figure2}(b). In the second regime, strong follower–leader attraction, $\mu_{fl}$, enables followers initially positioned within range of the leaders to move towards them. However, because in this regime, follower–follower attraction is comparatively weak, these migrating cells are unable to attract additional followers from the rear of the bulk, leaving the majority of the population behind. An example simulation in this regime is shown in the lower panels of Figure \ref{fig:figure2}(a) and corresponds to the region denoted by $\dag$ in the binary heatmap shown in Figure \ref{fig:figure2}(b). Figure \ref{fig:figure2}(b) shows a heatmap of $M_f/M_0$ for $\xi=50\,\mu\rm{m}$ with regions in which $M_f/M_0 = 0$ for large values of $\mu_{ff}$ relative to $\mu_{fl}$, and $M_f/M_0>0$ for smaller values of $\mu_{ff}$ relative to $\mu_{fl}$, corresponding to the two regimes described above, respectively. Adjacent to this heatmap is a binary heatmap clearly indicating the demarcation of these two regimes. 

As shown in Figure \ref{fig:figure3}(b), the fraction of transported follower mass increased monotonically with follower–leader attraction strength $\mu_{fl}$. However, for larger values of this parameter, this increase plateaued, indicating a fundamental limit on the proportion of followers that could be transported along the domain with leaders. The precise saturation level depended on the typical follower interaction range, $\xi$. For $\xi = 10\,\mu$m, only a very small fraction of followers, below $0.05$, migrated behind the leaders. For $\xi = 50\,\mu$m, the transported fraction did not exceed $0.2$. For $\xi = 100\,\mu$m, the transported fraction reached a maximum of approximately $0.35$, and consistently produced the largest mass of follower cells transported for a given set of interaction strengths ($\mu_{ff}, \mu_{fl})$ (Figure~\ref{fig:figure3}). Thus, for all parameter ranges considered, the majority of followers failed to migrate.
Here, we note that if $\xi$ is further increased, it is possible that a larger mass of follower cells could be transported along the domain. However, such values of $\xi$ would correspond to cell--cell communication over lengths that are far larger than the typical range of short-range communication or adhesion \textit{in vivo}.

Taken together, these results show that while the model of cell--cell adhesion considered here can generate small cohorts of followers trailing leaders, it cannot sustain extended streams akin to those observed \textit{in vivo}. Instead, follower transport is limited by the maximum length over which followers may interact with other cells. 

\section*{Discussion}

Our analysis demonstrates that, within the minimal non-local advection--diffusion framework considered here, short-range follower--follower and follower--leader adhesion are not sufficient to sustain spatially extended, motile follower streams. In the model, the extent of the follower population transported with the leaders remains constrained by the interaction lengthscale over which adhesive forces act. Thus, although the system can generate cohesive follower groups, it does not support the formation of long migrating streams under biologically plausible short-range interactions.

Numerical simulations of the coupled one-dimensional non-local PDE system consistently exhibit two distinct dynamical regimes, determined by the relative magnitudes of the attraction strengths $\mu_{ff}$ and $\mu_{fl}$. When follower--follower attraction is sufficiently strong, the follower population remains bound within its initial bulk, and leader attraction is unable to detach and transport it. By contrast, when follower--leader attraction is sufficiently strong, only those followers initially within the interaction range of leaders are advected forwards, while the remaining follower mass is left behind because follower--follower attraction is too weak to propagate motion through the bulk. In both cases, the transported follower mass is restricted by the finite range of the interaction kernel.

Furthermore, our results show that increasing follower--leader attraction does not remove this limitation. Instead, for fixed interaction lengthscales, $\xi$, the transported mass increases only up to a plateau, after which stronger attraction to leaders produces no substantial increase in the size of the migrating follower cohort. The saturation level depends on $\xi$, but for all values considered it remains bounded away from transport of the full follower population. This indicates that the inability to generate extended streams is not a consequence of parameter choice alone, but an intrinsic feature of the canonical short-range non-local adhesion framework.

As this study is based on numerical exploration of parameter space, we cannot exclude the existence of isolated parameter values for which the two competing effects balance more delicately. However, the absence of any broad parameter regime supporting extended follower transport suggests that such behaviour, if it exists, is not structurally robust within this model class. A key takeaway of the present work is that a widely used continuum formulation has an important limitation; short-range non-local adhesion and co-attraction can generate local cohesion, but do not by themselves produce long, coherent travelling streams.

This conclusion does not rule out adhesion as a biologically important process in the formation of follower cohorts. Rather, it suggests that something essential may be missing from how adhesion is represented in current continuum models. Additional mechanisms, such as anisotropic interactions, longer-range signalling, leaders responding to follower behaviour \citep{bernardi_painter_2025_leaders_followers}, or coupling to external fields or substrates, may also be required to maintain coordinated motion across large follower populations. Our results point to an open challenge: developing continuum models that can produce long, coherent travelling streams through purely mass-conserving, collective cell movement.

\section*{Acknowledgements}
The authors would like to thank Professor Raluca Eftimie for a valuable discussion leading to the conception of this study. T.J.J. is supported by funding from the Edmund J. Crampin Scholarship at Linacre College, University of Oxford; and the Engineering and Physical Sciences Research Council (EPSRC) (grant number EP/W524311/1). R.E.B. is supported by a grant from the Simons Foundation (MP-SIP-00001828). For the purpose of open access, the authors have applied a CC BY public copyright licence to any author accepted manuscript arising from this submission.

\bibliographystyle{unsrt}
\bibliography{BIBL}
\newpage
\appendix
\section*{Appendix A: Model Formulation and Numerical Scheme}

\subsubsection*{A.1 Numerical scheme}

We discretise the domain into $N=10^{4}$ mesh points, $\{x_i\}_{i=1}^N$ with uniform
spacing, $h=L/(N-1)$. The system is integrated explicitly in time using a
forward Euler scheme. The time step, $\Delta t$, is chosen conservatively to
satisfy a von Neumann stability constraint
\begin{equation}
\Delta t \;\leq\; \tfrac{h^2}{2D_{f}},
\label{hijun}
\end{equation}
with additional safety factor, $\theta\in(0,1)$ (here $\theta=0.05$), to ensure accurate results, giving a final time step of

\begin{equation}
\Delta t = \theta \frac{h^2}{2D_{f}}.
\end{equation}

In principle one may also impose an advection Courant--Friedrichs--Lewy
constraint arising from the taxis flux. The maximum taxis velocity scales
as (using $\|\nabla K_\xi\|_\infty \sim 1/\xi$ for unit-amplitude kernels)
\begin{equation}
  v_{\max} = \frac{(|\mu_{fl}| + |\mu_{ff}|)\,\kappa}{\xi},
\end{equation}
giving the condition
\begin{equation}
  \Delta t \leq \frac{h}{v_{\max}}.
\end{equation}
For the parameter ranges considered here
($\mu_{fl}, \mu_{ff} \in [10, 1000]\,\mu\mathrm{m}\,\mathrm{min}^{-1}$,
$\xi \in [10, 100]\,\mu\mathrm{m}$, and $\kappa = 1$),
the advection bound satisfies
$h / v_{\max} \geq 1 \times 10^{-3}\,\mathrm{min}$, which is comparable
in magnitude to the diffusion bound
$h^2 / (2 D_f) = 2 \times 10^{-3}\,\mathrm{min}$. Applying the safety
factor $\theta = 0.05$ to the diffusion bound in Equation~\eqref{hijun}
yields a time step of
$\Delta t = 1 \times 10^{-4}\,\mathrm{min}$, which satisfies both the
diffusion and advection constraints by an order of magnitude across all
simulations reported in this work.

\subsubsection*{A.1.1 Numerical leader update}

Leaders are transported at a speed $v_0\,\mu\rm{m}\,\text{min}^{-1}$ using a discrete shifting algorithm. The displacement increment per time step is
\begin{equation}
\delta s = \frac{v_0 \Delta t}{h}.
\end{equation}
Let $S$ denote the accumulated displacement in the $x$-direction. At each time step, $S$ is increased by $\delta s$, and whenever $\lfloor S \rfloor$ increases by one, the leader density profile, $\rho_l$, is shifted by one mesh point in the positive $x$-direction.

\subsubsection*{A.1.2 Numerical follower update}

Followers evolve via explicit forward Euler updates of the form
\begin{equation}
\rho_f^{n+1} = \rho_f^n
+ \Delta t\Big(D_{f}\Delta_h \rho_f^n
- \nabla_h \cdot J^n\Big),
\end{equation}
where superscripts denote time and $\Delta_h$ denotes a second-order finite-difference Laplacian with
homogeneous Neumann boundary conditions.

The taxis flux is written in conservative form as
\begin{equation}
J = \rho_fP\nabla C,
\qquad
P = \kappa - (\rho_l + \rho_f),
\end{equation}
where $C$ is the non-local attraction potential. Zero-flux boundary conditions are imposed on $\rho_{f}$ at $x=0\,\mu\rm{m}$ and $x=2000\,\mu\rm{m}$, such that the integrated follower mass remains constant throughout migration.

\subsubsection*{A.1.3 Discretisation of the taxis flux}

\noindent The divergence of the taxis flux is discretised using a finite-volume
formulation. Gradients of the attraction potential, $C$, are evaluated at cell
faces using centred differences,
\begin{equation}
(\partial_x C)_{i+\frac12}
= \frac{C_{i+1}-C_i}{h},
\end{equation}
and the corresponding face velocities are defined by
\begin{equation}
u_{i+\frac12}
= P_{i+\frac12}(\partial_x C)_{i+\frac12},
\end{equation}
where $i$ indexes the centres of mesh points and $P_{i+\frac12}$ is obtained by arithmetic averaging of $P$ between
adjacent mesh points.

The numerical flux through each cell face is computed using an upwind
discretisation,
\begin{equation}
J_{i+\frac12} =
\begin{cases}
\rho_{f,i}u_{i+\frac12}, & u_{i+\frac12} \ge 0, \\
\rho_{f,i+1}u_{i+\frac12}, & u_{i+\frac12} < 0,
\end{cases}
\end{equation}
which yields a conservative update and preserves non-negativity of the
follower density.

The discrete divergence of the flux is then given by
\begin{equation}
(\nabla_h \cdot J)_i
= \frac{J_{i+\frac12}-J_{i-\frac12}}{h}.
\end{equation}

\subsubsection*{A.1.4 Non-local convolution}

Given a kernel, $K$, and a discrete density, $Z$, the convolution is given by
\begin{equation}
(K * Z)(x_i) \;\approx\; h\sum_j K_{\xi}(x_i-x_j)Z(x_j),
\end{equation}
and is computed using a fast Fourier convolution with zero padding beyond the domain boundaries.

\subsubsection*{A.1.5 Follower cell diffusion constant}

In simulations, we choose a moderate diffusion constant of $D_{f} = 10\,\mu\rm{m}^2\rm{min}^{-1}$ for follower cells.
For one hour of migration, this results in a characteristic diffusive length-scale of 
\begin{equation}
\ell(60\,\rm{min}) = \sqrt{2 D_{\mathit{f}} t} = \sqrt{2 \times 10\,\mu\mathrm{m}^2\mathrm{min}^{-1} \times 60\,\mathrm{min}}
= \sqrt{1200}\,\mu\mathrm{m}
\approx 35\,\mu\mathrm{m}.
\end{equation}

Thus, with this choice of $D_{f}$, the expected spread of a cell over one hour is approximately 3.5 
cell lengths ($\sim 35\,\mu$m), though this diffusion is largely attenuated by the follower--follower attraction that is also present in the model. 

\subsection*{A.2 Initial conditions}

The initial condition consists of a follower bulk trailing a leader pulse
\begin{align}
\rho_{f}(x,0) &= \tfrac{f_0}{2}\left( \tanh\!\tfrac{x-x_a^f}{\delta}
- \tanh\!\tfrac{x-x_b^f}{\delta}\right), \\[4pt]
\rho_{l}(x,0) &= l_0\exp\!\left(-\tfrac{(x-x_0)^2}{2\sigma^2}\right),
\end{align}
with $0<x_a^f<x_b^f<x_0<L$, where $f_0=l_0\leq \kappa$. In simulations we take
$[x_a^f,x_b^f]=[65\,\mu\text{m},870\,\mu\text{m}]$, $x_0=895\,\mu\text{m}$, 
$\delta=5\,\mu\text{m}$, $\sigma=12\,\mu\text{m}$, and $f_0 = l_0 = 0.95$,
yielding a narrow leader peak located just ahead of an extended follower train (Figure \ref{fig:figure2}(a)).

\end{document}